\documentclass[12pt]{article}

\usepackage{newtxtext,newtxmath}

\usepackage{graphicx}

\usepackage[letterpaper,margin=1in]{geometry}

\renewenvironment{abstract}
	{\quotation}
	{\endquotation}

\date{}

\makeatletter
\renewcommand{\fnum@figure}{\textbf{Figure \thefigure}}
\renewcommand{\fnum@table}{\textbf{Table \thetable}}
\makeatother

\usepackage{scicite}

\usepackage{url}

\def\scititle{
    3D Optofluidic Control Using Reconfigurable Thermal Barriers
 
}
\title{\bfseries \boldmath \scititle}

\author{
		Falko~Schmidt$^{1\ast}$,
        Carlos~David~Gonz\'alez-G\'omez$^{2}$,
        Emilio~Ruiz-Reina$^{3}$,	
	    Ra\'ul~A.~Rica$^{2,4}$,\and
        Jaime~Ortega~Arroyo$^{1}$,\and Romain~Quidant$^{1\ast}$\and
	\small$^{1}$Department of Mechanical and Process Engineering, ETH Zurich, Zurich 8092, Switzerland.\and
	\small$^{2}$Universidad de Granada, Department of Applied Physics, Granada 18071, Spain.\and
 	\small$^{3}$Department of Applied Physics II, University of Malaga, Malaga 29071, Spain.\and
        \small$^{4}$Universidad de Granada, Research Unit `Modeling Nature' (MNat), Granada 18071, Spain.\and
	\small$^\ast$Corresponding authors. Emails: schmidtfa@ethz.ch, rquidant@ethz.ch
}

\begin{document} 

\maketitle

\begin{abstract} \bfseries \boldmath

\noindent Microfluidics has revolutionized control over small volumes through the use of physical barriers. However, the rigidity of these  barriers limits flexibility in applications.
We present an optofluidic toolbox that leverages structured light and photothermal conversion to create dynamic, reconfigurable fluidic boundaries.
This system enables precise manipulation of fluids and particles by generating 3D thermal landscapes with high spatial control. 
Our approach replicates the functions of traditional barriers while additionally allowing real-time reconfiguration for complex tasks, such as individual particle steering and size-based sorting in heterogeneous mixtures.
These results highlight the platform's potential for adaptive and multifunctional microfluidic systems in applications such as chemical synthesis, lab-on-chip devices, and microbiology, seamlessly integrating with existing setups due to its flexibility and minimal operation requirements.

\end{abstract}




\noindent Microfluidics
has become an indispensable tool for a wide range of applications in chemical and biological studies, including chemical synthesis, drug development, and single-cell studies.
Its high level of control over microenvironments has been achieved by utilizing physical barriers.
When strategically placed and shaped, these barriers enable comprehensive control over fluid manipulation, including pumping, mixing, splitting, merging, and confinement.
Additionally, they facilitate the isolation, trapping, and sorting of suspended analytes like colloidal particles, cells, and polymers \cite{nilsson2009}.\\
Common fabrication techniques for such barriers include injection molding, lithography, additive manufacturing, micromilling, and laser ablation \cite{attia2009,sugioka2014,bhattacharjee2016}.
While these methods offer high resolution in both fabrication and element positioning, they are typically optimized for specific target analytes and a fixed set of manipulation tasks.
This specificity reduces their ability to handle a broad spectrum of samples simultaneously, while the lack of tunability to adapt to varying analyte requirements further constrains their flexibility.
Moreover, fixed barriers cannot be dynamically reconfigured to suit different tasks nor allow real-time experimental decision-making.
Yet, both tunability and reconfigurability are critical, for instance when developing lab-on-chip devices, where multiple functionalities must be integrated to deliver a complete workflow environment.\\
Several strategies have been proposed to address these limitations, such as controlling flow paths in fixed pneumatic valve networks \cite{thorsen2002}, using electro-osmotic flow fields \cite{schasfoort1999}, magneto-hydrodynamics \cite{bau2003}, and hydrogels \cite{d2018, wada2020}, or creating virtual channels via hydrodynamics \cite{taylor2020} and thermo-rheological conversion \cite{krishnan2012}, among others highlighted in Ref.~\cite{paratore2022}.
In particular, virtual boundaries provide a flexible alternative to physical barriers by directly manipulating the fluid using body or surface forces.
However, many of these methods rely on external, bulky fluidic instrumentation, require compositional changes to the liquid, or suffer from low spatial resolution and slow response times. Currently, a broadly applicable solution is still missing that provides reconfigurability with single $\mu$m resolution at timescales well below 100~ms, operates without external fluid pumps, and works in standard aqueous solutions.\\
Here, we introduce an optofluidic toolbox that addresses these challenges by building reconfigurable virtual 
boundaries for particle manipulation, dynamically shaped using light.
We create virtual barriers by generating local heating via photothermal conversion.
Specifically, by using structured light to illuminate both surfaces of a microfluidic chamber, we generate controlled 3D patterns of heat that activate and drive fluid dynamics and particle motion throughout the volume.
Through rigorous engineering of this thermal landscape, we replicate the functions of physical barriers, enabling fluidic operations such as steering, splitting, and merging and particle manipulations such as trapping. 
Combining experiments and simulations, we study the interaction of colloidal particles with the surrounding fluid patterns using different types of optofluidic barriers.
Thanks to light-based activation and rapid temperature changes, our platform is fully reconfigurable, and demonstrates so, by dynamically switching the orientation of a steering barrier to control the direction of collective and individually addressed particles.
Finally, to show the potential of reconfigurable virtual barriers we explore their applicability to particle sorting by emulating an adaptive deterministic lateral displacement (DLD) assay that switches between particle collection and lateral separation of particles of varying sizes.

\begin{figure} 
	\centering
	\includegraphics[width=0.5\textwidth]{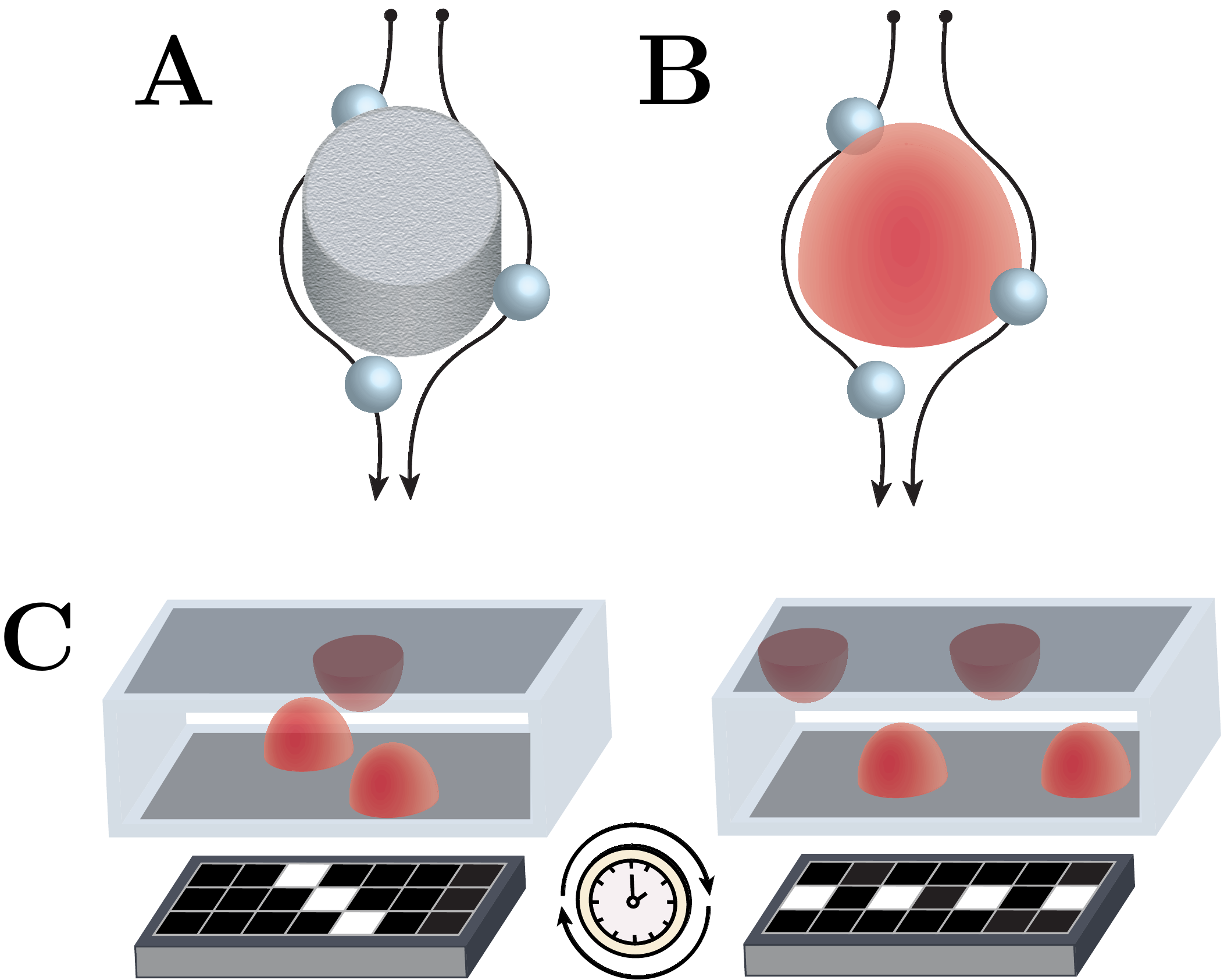} 
	\caption{\textbf{Concept of Reconfigurable Barriers.} Comparison of a physical pillar (\textbf{A}), and a virtual pillar (\textbf{B}) induced by a localized heat source, both altering the flow profile to direct particle movement around the obstacle. Reconfigurable barriers (\textbf{C}) are created using spatially modulated light (bottom), which generates localized heat sources through light absorption on the chamber surfaces (dark gray) in a microfluidic device (light gray). The size and shape of these optofluidic barriers can be dynamically adjusted over time by modifying the light patterns (white pixels).}
	\label{fig:1} 
\end{figure}

\subsection*{Generation of localized optofluidic barriers}
\begin{figure}
	\centering
	\includegraphics[width=\textwidth]{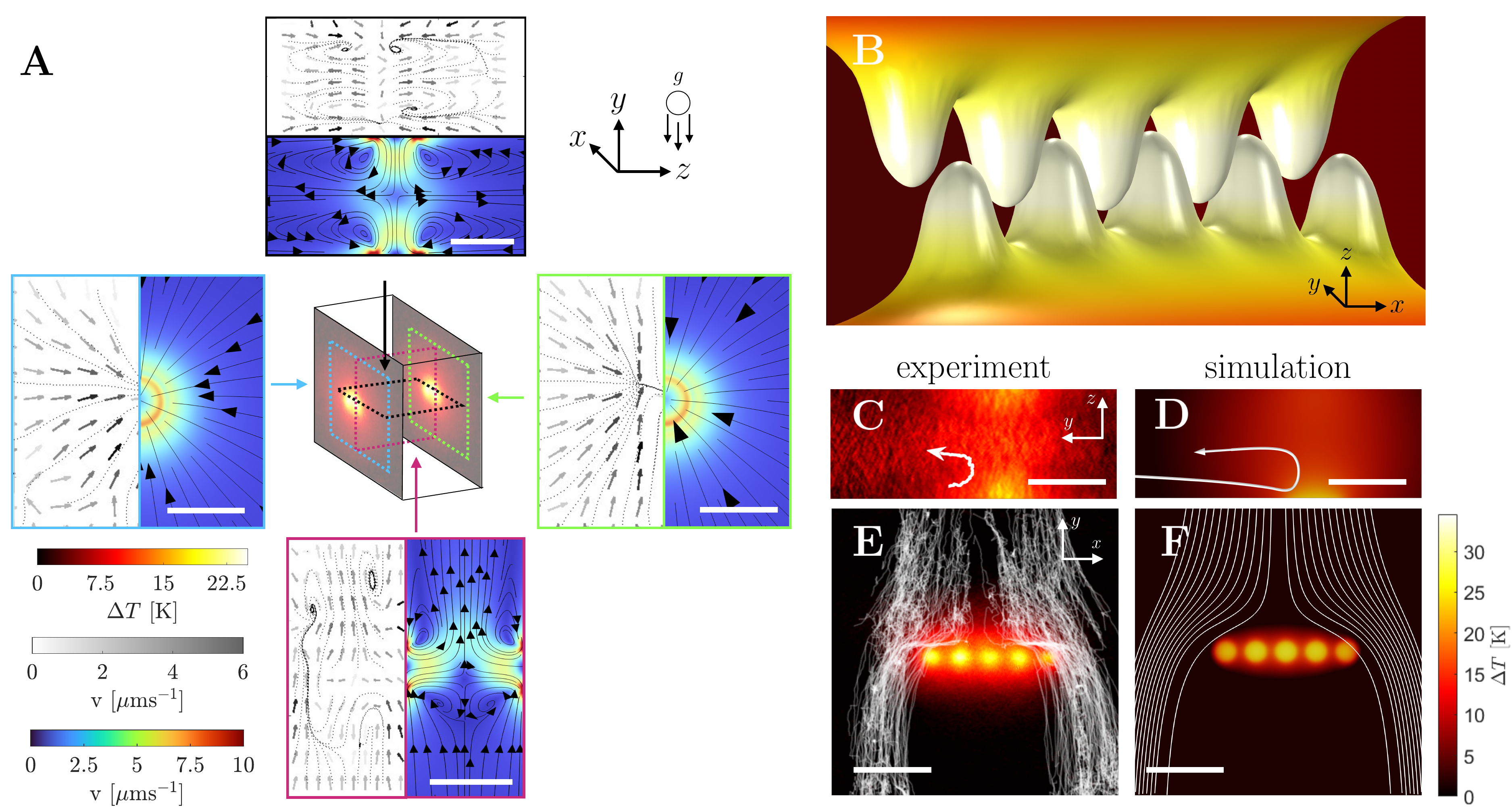}
	\caption{\textbf{Generation of a Localized Fluidic Barrier.}
		(\textbf{A}) Induced fluid flows within a microfluidic chamber (thickness $h=20~\mu$m) in experiments (white background) and simulations (blue background). A single laser beam generates a heat spot on the first (blue frame), and second absorbing surface (green frame), causing radially symmetric inward fluid flows due to thermo-osmosis near the surface ($z=2~\mu$m). These flows converge toward the center of the chamber from both sides, forming closed-loop flow profiles ($y=25~\mu$m, black frame). Along the vertical axis, additional convective flows push the fluid upwards ($x=25~\mu$m, magenta frame). Scale bars: $15~\mu$m.
(\textbf{B}) Simulated optofluidic barrier created by alternating heat spots.
(\textbf{C}) Side view from experiments and (\textbf{D}) simulations showing how a particle is deflected by the barrier. Scale bars: $5~\mu$m.
(\textbf{E}) Top view of experimentally measured particle trajectories, and (\textbf{F}) simulated trajectories, showing particles avoiding the barrier and continuing to sediment along the sides. Scale bars: $15~\mu$m.
}
	\label{fig:2}
\end{figure}

Physical pillars are key components in traditional barriers, distorting local fluid flows to control the directionality of particles and fluid alike (Fig.~\ref{fig:1}A).
To create virtual barriers, similar fluid flows must be induced by an external trigger, such as light.
Specifically, we exploit the light-to-heat conversion near an absorbing surface to disturb the local fluid flows and thus generate a virtual optofluidic pillar (Fig.~\ref{fig:1}B).
Spatially modulating the light impinging on the structure creates a reconfigurable optofluidic barrier made of individual pillars across the microfluidic chamber (Fig.~\ref{fig:1}C).\\
We demonstrate this principle in a microfluidic chamber (thickness $h=20~\mu$m) with surfaces coated in gold nanorods (AuNRs, absorption peak $\lambda_{\rm abs}=780$~nm), while remaining transparent for optical interrogation \cite{methods}.
When a laser resonantly excites the AuNRs ($\lambda_{\rm light}=807$~nm), both sides of the chamber heat up (Fig.~\ref{fig:2}A-D), in turn leading to thermally-induced fluid flows and particle motion (thermo-osmosis, convection, and thermophoresis).
To characterize the system, we measure the resulting temperature profiles and complex flow patterns using optical diffraction tomography (ODT) and digital holographic microscopy (DHM) \cite{methods}, respectively, and compare the results with simulations.
Near the surfaces, steep temperature differences up to 25~K induce thermo-osmotic flows \cite{franzl2022} directed radially inwards (Fig.~\ref{fig:2}A,B).
These inward flows converge towards the center of the chamber, where they meet and then redirect outwards, forming closed loops (Fig.~\ref{fig:2}C).
Additionally, convective flows, though small, are observed along the gravitational axis (Fig.~\ref{fig:2}D).
The strength of these flows depends on the chamber thickness, becoming predominant for $h>30~\mu$m (fig.~S1).\\
Since heat sources are additive, they 
can be combined in 3D to create a fluidic profile mimicking a physical barrier composed of individual pillars.
By alternating heat sources between surfaces an efficient optofluidic barrier with minimal gaps is created (Fig.~\ref{fig:2}E simulation).
The presence of double-sided heating is crucial, as particles would otherwise escape along the larger gaps in the temperature profile 
(fig.~S2). We assess the performance of our optofluidic barrier by positioning it in the path of a flux of sedimenting particles ($D=5~\mu$m) and count the number of particles passing through the barrier \cite{methods}.
As particles approach the barrier, temperature gradients near the heated surface, generating strong fluid flows, steer the particles toward the middle, and in some cases even repel them (Fig.~\ref{fig:2}F,G).
Particles then move parallel to the barrier at fixed distance before resuming their sedimentation-driven dynamics (Fig.~\ref{fig:2}H,I).
While thermo-osmotic and convective flows primarily govern the particle trajectories, thermophoresis significantly contributes as well (fig.~S3A).
Previous studies \cite{ciraulo2021,ruzzi2023} and control experiments (fig.~S3B,C) confirm that our dielectric particles (SiO$_2$, microparticles GmbH) are weakly thermophobic.
Consequently, these particles move away from high-temperature regions, enhancing the barrier's performance.
Slight discrepancies between experiment (Fig.~\ref{fig:2}H) and simulations (Fig.~\ref{fig:2}I) are caused by the challenge to measure the thermophoretic coefficient accurately \cite{piazza2008}.
Other effects, such as optical forces, are negligible here, as the laser beams used for heating were focused just outside the chamber (fig.~S3D). We further characterized the barrier’s efficiency by varying the heat source separation and laser power, which control the induced temperature increase.
At powers above 135~mW and distances below 13.3~$\mu$m, almost no particles pass through the optofluidic barrier, while lower powers and greater barrier separations reduce efficiency due to gaps in the temperature profile (fig.~S4).

\subsection*{Reconfigurability}
\begin{figure}
	\centering
	\includegraphics[width=0.8\textwidth]{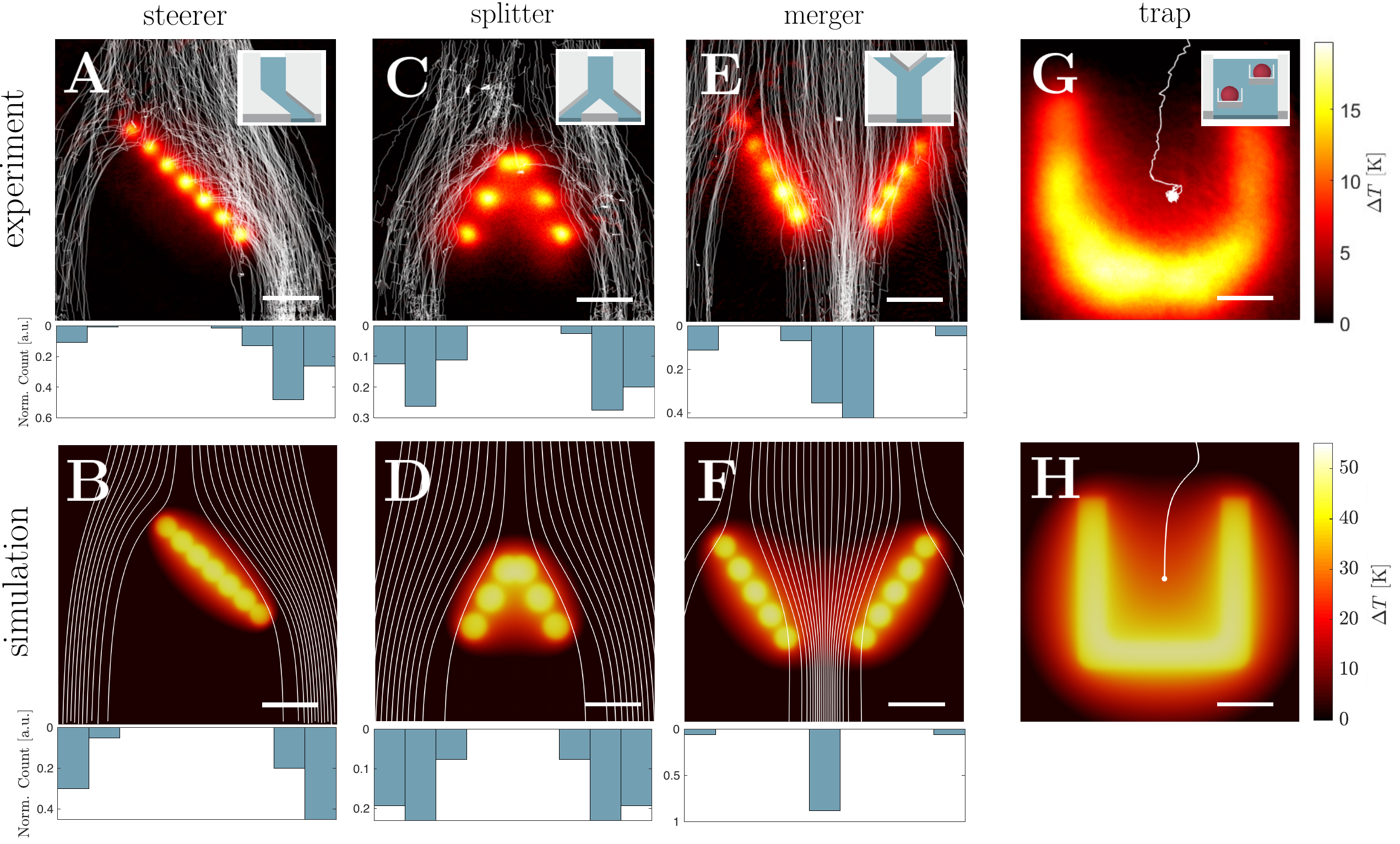}  
	\caption{\textbf{Types of Optofluidic Barriers.}
		(\textbf{A,C,E,G}) Experimental images of heat-induced barriers with particle trajectories (white), accompanied by 3D sketches of the equivalent microfluidic actuators (insets).
  (\textbf{B,D,F,H}) Simulated temperature profiles of the barriers with particle trajectories (white).
  (\textbf{A,B}) A steerer, created by a single barrier tilted at $-45^{\circ}$, deflects particles to the right, similar to the effect of a tilted microfluidic channel.
  (\textbf{C,D}) A splitter, created by two tilted barriers divides particle flow evenly into two, resembling a Y-shaped microfluidic channel with two outlets.
  (\textbf{E,F}) A merger directs particles to converge at the center, akin to a Y-channel with two inlets.
  (\textbf{G,H}) A U-shaped barrier traps a particle entering from the top, confining it from both sides to a single equilibrium point, similar to micropockets in traditional fluidic channels. Scale bars 10~$\mu$m.
}
	\label{fig:3}
\end{figure}

\begin{figure}
	\centering
	\includegraphics[width=\textwidth]{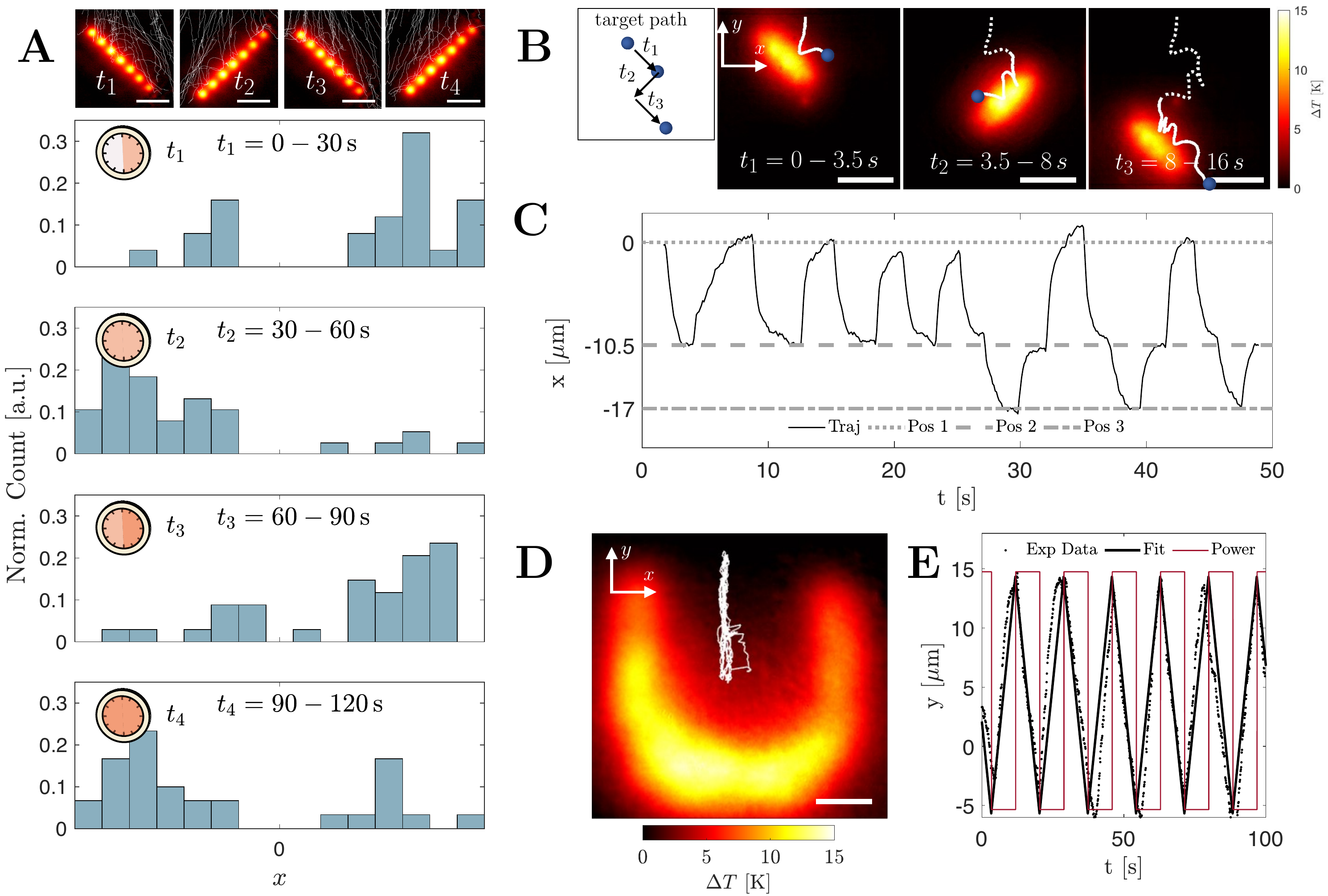}
	\caption{\textbf{Reconfigurability of Optofluidic Barriers.}
		(\textbf{A}) A single steerer barrier is periodically switched between two orientations ($-45^{\circ}$ and $45^{\circ}$) every 30~s, demonstrating that particles can be deflected left and right over time, as shown by the shifts in distribution in the histograms.
    (\textbf{B}) By defining a target path for a sedimenting particle, the steerer barrier is dynamically positioned and reoriented to guide the particle along the desired trajectory.
    (\textbf{C}) The trapping position of a particle is laterally shifted between three pre-defined positions by translating the U-shaped barrier, visible in the particle’s $x$-axis trajectory.
    (\textbf{D}) Vertical shifts in the trapped particle’s position are achieved by adjusting the laser input power.
    (\textbf{E}) A rectangular laser power signal (red) induces a triangular particle position response (black dots) as the particle transitions between two equilibrium points, with the position fitted by a triangular function with a 17~s period. Scale bar 10~$\mu$m.
}
	\label{fig:4}
\end{figure}

To demonstrate full reconfigurability in lab-on-chip devices, the utilized method must: (1) be able to set a desired function, and (2) execute changes in real-time \cite{paratore2022}.
We showcase both capabilities: the creation of barriers with diverse functionalities for particle manipulation (Fig.~\ref{fig:3}), and the dynamic switching of these configurations for both groups and individual particles, enabled by real-time feedback (Fig.~\ref{fig:4}).
We achieve these functionalities by engineering the thermal landscape through projected light patterns that generate optofluidic barriers of varying length, orientation, and composition.
We focused our particle and fluid characterization on three basic barrier types, which mimic the core functions of classical microfluidics (Fig.~\ref{fig:3}A,C,E,G, insets): steerer (Fig.~\ref{fig:3}A,B), splitter (Fig.~\ref{fig:3}C,D), and merger (Fig.~\ref{fig:3}E,F).
These important fundamental functions enable precise control over the motion of suspended particles as integral parts of larger workflows.\\
For the steerer, we experimentally created a -45$^\circ$ tilted barrier that redirected particles towards the right (Fig.~\ref{fig:3}A).
This optofluidic barrier achieves the same functionality as a classical microfluidic design using a single inclined channel (Fig.~\ref{fig:3}A, inset).
Simulations of such optofluidic barrier show similar steering behavior, with a higher concentration of particles emerging on the right (Fig.~\ref{fig:3}B, histogram).
While most particles moved rightwards, thermophoresis drove those far left of the barrier further left, providing an additional tunable parameter compared to physical barriers via particle surface functionalization \cite{ruzzi2023}.
By symmetry, a 45$^{\circ}$ tilted barrier steers particles to the left (Fig.~\ref{fig:4}A).
Similarly, a steerer can be built using predominant convective patterns in microfluidic chambers of $h\geq50~\mu$m which particles tend to follow (fig.~S5).\\
To create a splitter, we combined two tilted barriers, thus dividing the particle flow evenly between the left and right sections (Fig.~\ref{fig:3}C,D).
This design resembles a Y-channel microfluidic architecture made of one input and two output channels (Fig.~\ref{fig:3}C, inset).
Such functionality creates two distinct regions of increased particle concentration while leaving a void in the center.
The slight asymmetry observed in experiments is attributed to the limited number of recorded trajectories (Fig.~\ref{fig:3}C, histogram).\\
Conversely, we created a merger by inverting the splitter design, leaving a gap of $20~\mu$m at the bottom to focus particles from both sides toward the center (Fig.~\ref{fig:3}E,F), mimicking a Y-channel with two inlets and one outlet (Fig.~\ref{fig:3}E, inset).
Beyond these linear barrier designs, we further demonstrate that more complex patterns offer additional functionalities than particle steering, such as trapping.
Specifically a U-shaped barrier allows particles to enter from the top, while confining them from all other directions (Fig.~\ref{fig:3}G,H).
Such functionality is akin to micropockets in flow channels (Fig.~3D inset) that allow, for instance, long-term studies of trapped analytes \cite{krishnan2010,takahashi2018}.\\
This specific example demonstrates that more complex workflows for fluid and particle actuation can be delivered by the superposition of multiple barrier types (fig.~S6).
Optofluidic barriers not only resemble the functionalities of structural barriers but offer additional degrees of freedom via the interaction of temperature gradients with fluids (thermo-osmosis, convection) and particles (thermophoresis) separately.
The combination of these individually tunable thermal effects enables a more precise control over particle motion, concentration, and targeted positioning such as for enhancing analyte detection and synthesis reactions.\\

\noindent Since temperature gradients on the microscale are induced within milliseconds \cite{vasista2024}, these optofluidic barriers can be reconfigured in real time to meet changing experimental conditions (Fig.~\ref{fig:4}), offering significant advantages over static physical barriers.
Using an liquid-crystal-on-silicon (LCOS)-based spatial light modulator (60~Hz refresh rate \cite{methods}), we dynamically switch between optofluidic barriers. 
As an example demonstrating passive reconfigurability, we periodically flipped a steerer barrier, altering the spatial distribution of particles from right to left (Fig.~\ref{fig:4}A).
For statistical accuracy each steerer orientation was maintained for 30 seconds; however, much faster switching is possible, as shown by the targeted steering of a single particle (Fig.~\ref{fig:4}B).
To achieve active feedback at the single-particle level, we defined a target path and projected a barrier to induce the desired response based on real-time particle tracking \cite{methods}.
For example, a steerer with a -45$^\circ$ tilt directs the particle downward and to the right, while the opposite tilt moves the particle left and down.
This process strongly depends on the particle's individual interaction with the barrier, as its exact position cannot be precisely predicted in the presence of random Brownian motion. Another example of time-dependent single particle manipulation was demonstrated by shifting the equilibrium position of an optofluidic trap (Fig.~\ref{fig:4}C-E).
Here, shifting the position of the U-shaped barrier tuned the particle's lateral position (Fig.~\ref{fig:4}C), while adjusting the input laser power provided vertical control (Fig.~\ref{fig:4}D,E).
The latter example illustrates reconfigurability beyond spatially reshaping the thermal environment; instead, the particle's motion is controlled by tuning the convective flow contribution, which is directly determined by the amount of power converted to heat (Fig.~\ref{fig:4}E). Altogether these results demonstrate that optofluidic barriers not only emulate functionalities associated to conventional barriers, but also deliver added features such as feedback-based positional control thanks to the tunability and reconfigurability of thermal energy landscapes.

\subsection*{Particle sorting}
\begin{figure}
	\centering
	\includegraphics[width=0.8\textwidth]{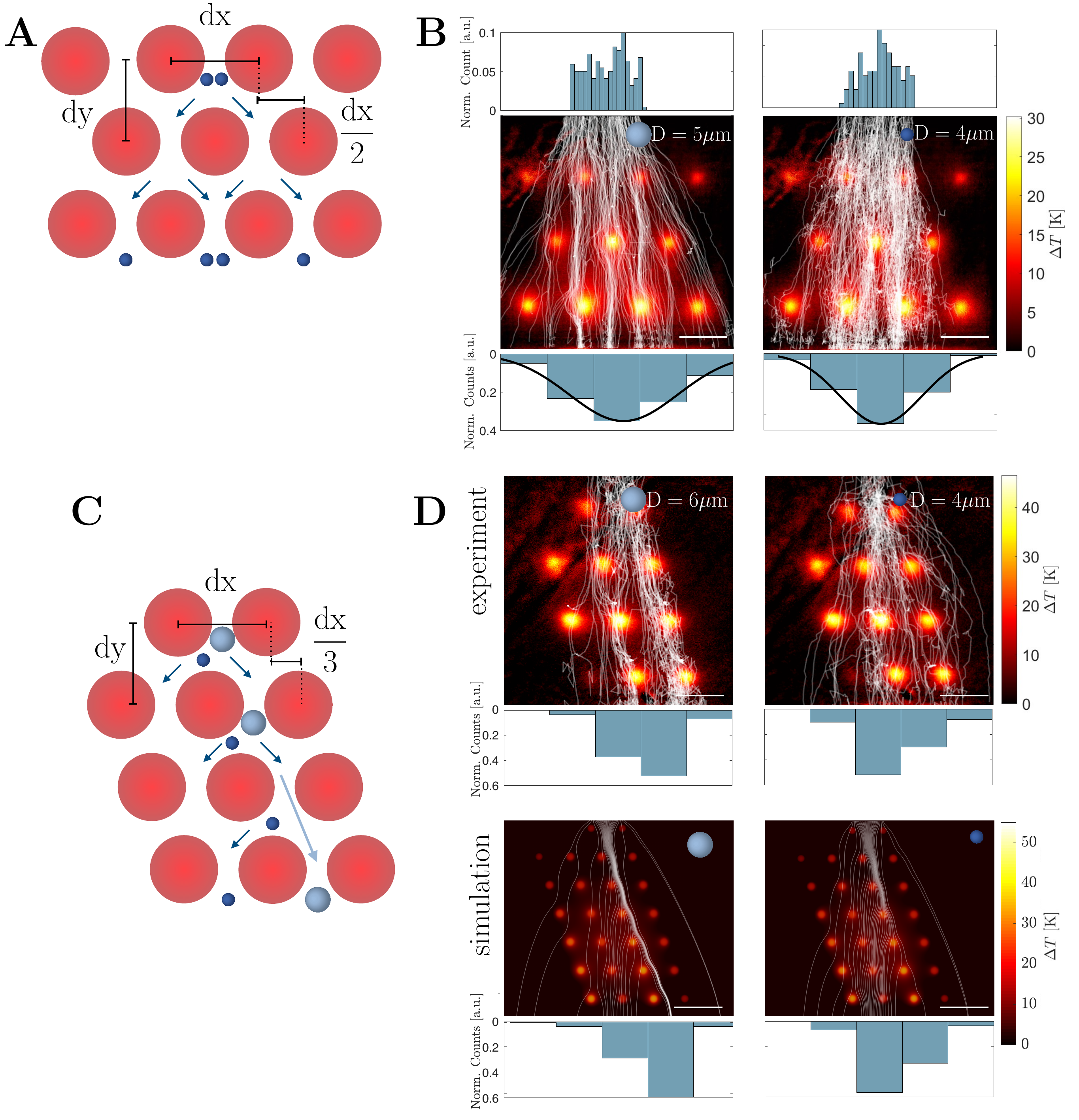}
	\caption{\textbf{Optofluidic Particle Sorting.}
    (\textbf{A}) A periodic arrangement of optofluidic pillars mimics the fluidic equivalent of the Galton board experiment.
    (\textbf{B}) Particles move from top to bottom through the periodic optofluidic pattern with an irregular starting distribution, resulting in a binomial distribution of particle positions at the bottom (after normalization), which converges to a normal distribution (black line) for many trials ($>$200). This behavior is independent of particle size, demonstrated here for diameters $D=5~\mu$m ($dx=27~\mu$m, $dy=30~\mu$m) and $D=4~\mu$m ($dx=24~\mu$m, $dy=27~\mu$m).
    (\textbf{C}) By laterally shifting consecutive rows of barriers by $dx/3$, the pattern replicates deterministic lateral displacement experiments for particle sorting.
    (\textbf{D}) Larger particles ($D=6~\mu$m) are displaced up to 60\% toward the right-bottom segment, while smaller particles ($D=4~\mu$m) are less deflected (31\%), effectively separating the two species. This behavior is observed in both experiments and simulations. Scale bars experiment $10~\mu$m, simulations $20~\mu$m. 
    }
	\label{fig:5}
\end{figure}

To demonstrate the potential of reconfigurable optofluidic barriers for advanced particle manipulation, we investigated their performance in the context of particle sorting.
By tuning the size and spacing of the fluidic pillars, we enhanced the sorting efficiency for particles of various sizes.
Initially, we arranged the pillars in a rhombic lattice, with each row shifted by half the interpillar distance, $dx/2$, allowing free adjustment of the spacing along $x$ and $y$ (Fig.~\ref{fig:5}A).
This pillar configuration, analogous to a microscopic Galton board, not only physically demonstrates the central limit theorem, but also showcases how fluidic barriers guide particles into distinct distributions.\\
When particles pass through the grid, each encounter with a pillar results in a 50\% chance of either a left or right displacement, which after several rows results in a binomial particle distribution.
Although our field of view is limited by the number of rows, we have experimentally verified the central limit theorem, as the histogram approximates a normal distribution after tracking only 200 individual particles starting with an irregular distribution (Fig.~\ref{fig:5}B).
We further confirmed the method's versatility for different particle sizes by reconfiguring the pillar spacing accordingly (e.g., $dx=27~\mu$m, $dy=30~\mu$m for $D=4~\mu$m, and $dx=24~\mu$m, $dy=27~\mu$m for $D=5~\mu$m and $D=4~\mu$m).\\
Building on this optofluidic array, we demonstrated particle size sorting using a modified deterministic lateral displacement (DLD) assay \cite{huang2004}.
DLD assays separate particles by size by applying a slight shift between consecutive rows of pillars ($dx/3$), which displaces larger particles along a different path compared to smaller ones (Fig.~\ref{fig:5}C).
In the current configuration, the pillars deflect larger particles to the right consistently following the gaps between pillars.
Smaller particles, however, follow a different path, with a leftward bias after passing several rows, effectively separating them from the larger particles.
Experimentally we confirmed particle sorting for a sample composed of $D=6.2~\mu$m and $D=4.4~\mu$m particles using an optofluidic pillar pattern with the following parameters $dx=17.5~\mu$m and $dy=20~\mu$m. 
Fig.~\ref{fig:5}C shows a distinct change in distribution when comparing both sizes.
Larger particles had a higher tendency ($\approx60\%$) to exit through the right channel, while smaller particles predominantly exited through the left ($\approx69\%$) well in agreement with the DLD design (Fig.~\ref{fig:5}D, fig.~S7 for $D=2.8$ and $1.5~\mu$m).
For particles below a threshold size ($D \approx 3.5\mu$m), the separation mechanism shifts as the balance between convection and sedimentation changes.
While smaller particles move upwards due to a greater convective contribution, larger ones are driven downwards by sedimentation. 
Therefore, samples with a mixture of particles with sizes above and below the threshold naturally separate, while for mixtures with sizes entirely below or above said threshold, the DLD pattern is realigned along the general direction of particle motion (Fig.~S7).
Though our experiments were not optimized for high separation efficiency, we achieved effective sorting within a working length of $<100~\mu$m, significantly shorter than the millimeter-scale devices used in conventional DLD systems \cite{huang2004}.
We used simulations to first confirm their agreement with experiments and then extrapolated the system beyond the experimental field of view (Fig.~\ref{fig:5}D).
We observed that smaller particles deviate left just before the third row, while larger ones continue downstream.
Adjusting the temperature profile within the DLD array improved sorting efficiency by exploiting thermophobic behavior, guiding large particles along the designed gap in the pattern, and enhancing the lateral displacement of smaller ones (fig.~S9). To better understand how the thermal landscape of the DLD assay affects particle deflection depending on their size, we have performed a sweep for the pillar spacing $dx$ and $dy$ and found a distinct regime in which only the larger size is deflected (fig.~S8).
This highlights the importance of thermophoretic effects for sorting, providing additional criteria beyond particle size (Supplementary Text).\\
These results demonstrate that combining thermally tunable fields with optofluidic barriers improves sorting precision and efficiency, particularly for closely sized particle mixtures, thereby providing a significant advantage over static DLD designs for compact microfluidic systems.\\

\noindent In conclusion, we have demonstrated that light-induced heating within a microfluidic chamber can create dynamically tunable optofluidic obstacles.
Our results show that these fluidic barriers not only manipulate particle flow similarly to physical ones, but can be further engineered to perform more complex microfluidic tasks such as steering, merging, splitting, and trapping in a reconfigurable manner.
The reconfigurability of these optofluidic barriers allows for virtual integration of multiple tasks within a single microenvironment and real-time adaptation to a broad range of analytes, including particles and potentially cells, though further studies are needed to assess cell viability \cite{ciraulo2021,ding2022}.\\
While our platform is based on thermally induced fluid flows to generate virtual boundaries, the principle of using spatial light modulation to induce localized temperature changes can be extended to other temperature-dependent processes \cite{chen2022}.
For instance, sol-gel phase transitions in polymers could be exploited to create reversible physical boundaries \cite{d2018, wada2020}, or density fluctuations of a critical binary mixture can induce surface forces for the nanopositioning and orientation of microelectromechanical systems \cite{schmidt2023,wang2024}.
Alternatively, optofluidic barriers could be induced via other light-activated processes beyond heating, for instance, light-induced changes in the rheological properties of liquids \cite{krishnan2012}.\\
While scalability and high throughput remain challenges for industrial integration, alternative devices like digital micromirror arrays or acousto-optical modulators could offer higher frequency actuation, overcoming current limitations.
The rapid heat induction enables real-time adaptation with feedback loops, making our optofluidic platform a versatile tool for reconfigurable microfluidics.



\clearpage 

%
\bibliography{OF3D_biblio} 
\bibliographystyle{sciencemag}

%
%
%
%
%
%


\section*{Acknowledgments}
\paragraph*{Funding:}

FS acknowledges funding from the SNSF PostDoctoral Fellowship (TMPFP2 209765). 
CDGG and RAR: Grant PID2021-127427NB-I00 funded by MICIU/AEI/
10.13039/501100011033 and, by ERDF A way of making Europe.

\paragraph*{Author contributions:}
\paragraph*{Competing interests:}
``There are no competing interests to declare.''




\newpage



\end{document}